\documentclass[aps,twocolumn,amssymb]{revtex4}
\usepackage{amssymb}
\usepackage{graphicx}
\usepackage{color}
\usepackage{textcomp}
\usepackage{ulem}
\usepackage{lipsum}
\usepackage{amsmath}

\begin{document}

\title{Fourth order correlation of baryon number and electric charge as a better magnetometer of QCD}
\author{Shijun Mao$^1$}
 \email{maoshijun@mail.xjtu.edu.cn}
 \author{Shuai Yang$^1$, Sicheng Lin$^1$, Xinran Yang$^1$, Guoyun Shao$^1$, Wen-Chao Zhang$^2$}
\affiliation{$^1$ School of Physics, Xi'an Jiaotong University, Xi'an, Shaanxi 710049, China\\
$^2$ School of Physics and Information Technology, Shaanxi Normal University, Xi'an 710119, China}

\begin{abstract}
This work focuses on the fourth order correlations $\chi^{BQ}_{31}$, $\chi^{QB}_{31}$, $\chi^{BQ}_{22}$, $\chi^{BS}_{31}$, $\chi^{SB}_{31}$, $\chi^{BS}_{22}$, $\chi^{QS}_{31}$, $\chi^{SQ}_{31}$, $\chi^{QS}_{22}$, $\chi^{BQS}_{211}$, $\chi^{QBS}_{211}$, $\chi^{SBQ}_{211}$ of baryon number $B$, electric charge $Q$ and strangeness $S$ at finite temperature, magnetic field and vanishing quark chemical potential. The study is carried out in the framework of a three-flavor PNJL model, considering both cases with and without inverse magnetic catalysis effect. We find that, fourth order correlations $\chi^{BQ}_{31}$ at chiral restoration phase transition is more sensitive to the magnetic field than other second order and fourth order correlations and fluctuations, and can be served as a more effective magnetometer of QCD.
\end{abstract}

\date{\today}

\maketitle

\section{Introduction}
In recent years, the study on Quantum Chromodynamics (QCD) phase structure under external electromagnetic fields has attracted much attention~\cite{review0,review1,review2,review3,review4,review5,lattice1,lattice2,lattice2sep,lattice4,lattice5,lattice6,lattice7,lattice9,ding2025,lattice8,fukushima,mao,maosep1,maosep2,maosep3,maosep4,kamikado,bf1,bf13,bf2,bf3,bf5,bf51,bf52,bf8,bf9,bf11,db1,db1sep,db2,db3,db5,db6,pnjl1,pnjl1sep,pnjl2,pnjl3,pnjl4,pqm,ferr1,ferr2,mhuang,meimao1,t0effect,meihuangmao,t0effectmao,efield1,efield2,efield3,efield4,addding,ding2022,dingprl2024,dingref31,dingref33,dingref34,dingref35,dingref36,mao2pnjl,maoyangprd,zhangwc}. Since strong magnetic field exists in the core of compact stars and in the initial stage of relativistic heavy ion collisions~\cite{b0,b1,b2,b3,b4}. An apparent effect of external magnetic field on QCD phase transitions is to change the strength of phase transitions, which leads to an increasing phase transition strength with increasing magnetic field~\cite{lattice1,lattice2,db5,pqm,dingref35,maoyangprd,mao2pnjl}.

The thermodynamical properties of QCD matter are influenced by the external magnetic field. Among the thermodynamic quantities, the correlations and fluctuations of the conserved charges are accessible both in theoretical calculations and experimental measurements and they can serve as the useful probes to study the QCD phase transitions~\cite{xwork24,xwork25,xwork26,xwork27,xwork28,xwork29,xwork30,ding99,ding100,shaoyang2025}. However, the correlations and fluctuations are much less explored at the chiral restoration and deconfinement phase transitions with finite temperature, magnetic field and vanishing quark chemical potential. Except for the  Lattice QCD (LQCD) calculations~\cite{addding,ding2022,dingprl2024,ding2025}, the analytical investigations at vanishing quark chemical potential, finite temperature and magnetic field~\cite{dingref31,dingref33,dingref34,dingref35,dingref36,mao2pnjl,maoyangprd,zhangwc} have been conducted in frame of the hadron resonance gas model, Polyakov loop extended Nambu-Jona-Lasinio (PNJL) model and Polyakov loop extended chiral quark model. The second order correlation between baryon number and electric charge is predicted to served as the possible magnetometer of QCD (probing the presence of magnetic fields in QCD matter)~\cite{dingref31,addding,ding2022,dingprl2024,ding2025,mao2pnjl,maoyangprd}. Higher order correlations between conserved charges are not considered under external magnetic field.

The LQCD simulations~\cite{lattice1,lattice2,lattice2sep,lattice4,lattice5,lattice6,lattice7,lattice9,ding2025} observe the inverse magnetic catalysis (IMC) phenomena in the chiral restoration phase transition, which means the decreasing chiral condensates of $u/d$ quarks near the pseudo-critical temperature $T^c_{pc}$ of chiral restoration phase transition and the decreasing pseudo-critical temperature $T^c_{pc}$ under external magnetic field. Meanwhile, it is reported that the renormalized Polyakov loop increases with magnetic fields and the transition temperature $T^d_{pc}$ of deconfinement decreases as the magnetic field grows~\cite{lattice1,lattice2,lattice2sep,lattice4,lattice5,lattice9}. On analytical side, how to explain the inverse magnetic catalysis phenomena is still an open question. Many scenarios are proposed~\cite{fukushima,mao,maosep1,maosep2,maosep3,maosep4,kamikado,bf1,lattice9,bf13,bf2,bf3,bf5,bf51,bf52,bf8,bf9,bf11,db1,db1sep,db2,db3,db5,db6,pnjl1,pnjl1sep,pnjl2,pnjl3,pnjl4,pqm,ferr1,ferr2,mhuang,meimao1,t0effect,meihuangmao,t0effectmao}, such as magnetic inhibition of mesons, sphalerons, gluon screening effect, weakening of strong coupling, and anomalous magnetic moment.

In this work, we focus on the fourth order correlations $\chi^{BQ}_{31}$, $\chi^{QB}_{31}$, $\chi^{BQ}_{22}$, $\chi^{BS}_{31}$, $\chi^{SB}_{31}$, $\chi^{BS}_{22}$, $\chi^{QS}_{31}$, $\chi^{SQ}_{31}$, $\chi^{QS}_{22}$, $\chi^{BQS}_{211}$, $\chi^{QBS}_{211}$, $\chi^{SBQ}_{211}$ of baryon number $B$, electric charge $Q$ and strangeness $S$ at finite temperature, magnetic field and vanishing quark chemical potential. The study is carried out in frame of a three-flavor PNJL model. The IMC effect is introduced into the PNJL model through the magnetic field dependent coupling between quarks~\cite{bf8,bf9,maoyangprd,mao2pnjl,geb1,su3meson4,mao11,tian} and magnetic field dependent interaction between quarks and Polyakov loop~\cite{t0effectmao,t0effect,pnjl3,maoyangprd,mao2pnjl}, respectively. The comparison between the results in the cases with and without IMC effect is made. We find that, fourth order correlations $\chi^{BQ}_{31}$ at chiral restoration process is more sensitive to the magnetic field than other second order and fourth order correlations and fluctuations, and can be served as a better magnetometer of QCD.

The paper is organized as follows. After the brief introduction, Section~\ref{2fframe} presents our three-flavor magnetized PNJL model and the definition of correlations and fluctuations of baryon number $B$, electric charge $Q$ and strangeness $S$ at finite temperature, magnetic field and vanishing quark chemical potential. Section~\ref{results} discusses the numerical results of fourth order correlations in the cases without IMC effect and with IMC effect. Finally, we give the summary in Sec.\ref{summary}.

\section{theoretical framework}
\label{2fframe}

The three-flavor PNJL model under external magnetic field is defined with the Lagrangian density~\cite{pnjl5,pnjl6,pnjl7,pnjl8,pnjl9,pnjl10,pnjl12},
\begin{eqnarray}
\mathcal{L}&=&\bar{\psi}\left(i\gamma^{\mu}D_{\mu}-\hat{m}_0\right)\psi+\mathcal{L}_{ 4}+\mathcal{L}_{6}-{\cal U}(\Phi,\bar\Phi),\\
\mathcal{L}_{ 4}&=&G\sum_{\alpha=0}^{8}\left[(\bar{\psi}\lambda^{\alpha}\psi)^2+(\bar{\psi}i\gamma_5\lambda^{\alpha}\psi)^2\right],\nonumber \\
\mathcal{L}_{6}&=&-K\left[\text{det}\bar{\psi}(1+\gamma_5)\psi+\text{det}\bar{\psi}(1-\gamma_5)\psi \right],\nonumber\\
{\cal U}(\Phi,{\bar \Phi}) &=&  T^4 \left[-{b_2(t)\over 2} \bar\Phi\Phi -{b_3\over 6}\left({\bar\Phi}^3+\Phi^3\right)+{b_4\over 4}\left(\bar\Phi\Phi\right)^2\right].\nonumber
\label{lagrangian}
\end{eqnarray}
The covariant derivative $D^\mu=\partial^\mu+i Q A^\mu-i {\cal A}^\mu$ couples quarks to the two external fields, the magnetic field ${\bf B}=\nabla\times{\bf A}$ and the temporal gluon field  ${\cal A}^\mu=\delta^\mu_0 {\cal A}^0$ with ${\cal A}^0=g{\cal A}^0_a \lambda_a/2=-i{\cal A}_4$ in Euclidean space. The gauge coupling $g$ is combined with the SU(3) gauge field ${\cal A}^0_a(x)$ to define ${\cal A}^\mu(x)$, and $\lambda_a$ are the Gell-Mann matrices in color space. We consider magnetic field ${\bf B}=(0, 0, B)$ along the $z$-axis by setting $A_\mu=(0,0,x B,0)$ in Landau gauge, which couples quarks of electric charge $Q=\text{diag}(Q_u,Q_d,Q_s)=\text{diag}(2/3 e,-1/3 e,-1/3 e)$ and current mass $\hat{m}_0=\text{diag}(m^u_0,m_0^d,m_0^s)$. The four-fermion interaction $\mathcal{L}_{4}$ represents the interaction in scalar and pseudo-scalar channels, with coupling constant $G$ and Gell-Mann matrices $\lambda^{\alpha},\ \alpha=1,2,...,8$ and $\lambda^0=\sqrt{2/3} \mathbf{I}$ in flavor space. The six-fermion interaction or Kobayashi-Maskawa-'t Hooft term $\mathcal{L}_{6}$ with coupling constant $K$ is related to the $U_A(1)$ anomaly~\cite{tHooft1,tHooft2,tHooft3,tHooft4,tHooft5}. The Polyakov potential ${\cal U}(\Phi,\bar\Phi)$ describes deconfinement, where $\Phi$ is the trace of the Polyakov loop $\Phi=\left({\text {Tr}}_c L \right)/N_c$, with $L({\bf x})={\cal P} \text {exp}[i \int^\beta_0 d \tau {\cal A}_4({\bf x},\tau)]= \text {exp}[i \beta {\cal A}_4 ]$ and $\beta=1/T$, the coefficient $b_2(t)=b_{20}+b_{21} t+b_{22} t^2+b_{23} t^3$ with $t=T_0/T$ is temperature dependent, and the other coefficients $b_3$ and $b_4$ are constants. Here, $T_0$ represents the critical temperature of deconfinement phase transition in pure gauge case.

By converting the six-fermion interaction into an effective four-fermion interaction in the mean field approximation, the Lagrangian density can be rewritten as~\cite{3njlrehberg,3pnjlmei}
	\begin{eqnarray}
		\mathcal{L}&=&\bar{\psi}\left(i\gamma^{\mu}D_{\mu}-\hat{m}_0\right)\psi-{\cal U}(\Phi,\bar\Phi) \\
		&+&\sum_{\alpha=0}^{8}\left[K_\alpha^-\left(\bar{\psi}\lambda^\alpha\psi\right)^2+K_\alpha^+\left(\bar{\psi}i\gamma_5\lambda^\alpha\psi\right)^2\right]\nonumber\\	&+&K_{30}^-\left(\bar{\psi}\lambda^3\psi\right)\left(\bar{\psi}\lambda^0\psi\right)+K_{30}^+\left(\bar{\psi}i\gamma_5\lambda^3\psi\right)\left(\bar{\psi}i\gamma_5\lambda^0\psi\right)\nonumber\\	&+&K_{03}^-\left(\bar{\psi}\lambda^0\psi\right)\left(\bar{\psi}\lambda^3\psi\right)+K_{03}^+\left(\bar{\psi}i\gamma_5\lambda^0\psi\right)\left(\bar{\psi}i\gamma_5\lambda^3\psi\right)\nonumber\\	&+&K_{80}^-\left(\bar{\psi}\lambda^8\psi\right)\left(\bar{\psi}\lambda^0\psi\right)+K_{80}^+\left(\bar{\psi}i\gamma_5\lambda^8\psi\right)\left(\bar{\psi}i\gamma_5\lambda^0\psi\right)\nonumber\\	&+&K_{08}^-\left(\bar{\psi}\lambda^0\psi\right)\left(\bar{\psi}\lambda^8\psi\right)+K_{08}^+\left(\bar{\psi}i\gamma_5\lambda^0\psi\right)\left(\bar{\psi}i\gamma_5\lambda^8\psi\right)\nonumber\\	&+&K_{83}^-\left(\bar{\psi}\lambda^8\psi\right)\left(\bar{\psi}\lambda^3\psi\right)+K_{83}^+\left(\bar{\psi}i\gamma_5\lambda^8\psi\right)\left(\bar{\psi}i\gamma_5\lambda^3\psi\right)\nonumber\\	&+&K_{38}^-\left(\bar{\psi}\lambda^3\psi\right)\left(\bar{\psi}\lambda^8\psi\right)+K_{38}^+\left(\bar{\psi}i\gamma_5\lambda^3\psi\right)\left(\bar{\psi}i\gamma_5\lambda^8\psi\right)\nonumber,
		\label{semilagrangian}
	\end{eqnarray}
	with the effective coupling constants
	\begin{eqnarray}
		\label{constants}
		&&K_0^\pm=G\pm\frac{1}{3}K\left(\sigma_u+\sigma_d+\sigma_s\right),\\
		&&K_1^\pm=K_2^\pm=K_3^\pm=G\mp\frac{1}{2}K\sigma_s,\nonumber\\
		&&K_4^\pm=K_5^\pm=G\mp\frac{1}{2}K\sigma_d,\nonumber\\
		&&K_6^\pm=K_7^\pm=G\mp\frac{1}{2}K\sigma_u,\nonumber\\
		&&K_8^\pm=G\mp\frac{1}{6}K\left(2\sigma_u+2\sigma_d-\sigma_s\right),\nonumber\\
		&&K_{03}^\pm=K_{30}^\pm=\mp\frac{1}{2\sqrt{6}}K\left(\sigma_u-\sigma_d\right),\nonumber\\
		&&K_{08}^\pm=K_{80}^\pm=\mp\frac{\sqrt{2}}{12}K\left(\sigma_u+\sigma_d-2\sigma_s\right),\nonumber\\
		&&K_{38}^\pm=K_{83}^\pm=\pm\frac{1}{2\sqrt{3}}K\left(\sigma_u-\sigma_d\right),	\nonumber
	\end{eqnarray}
	and chiral condensates
	\begin{eqnarray}
		\sigma_u=\langle\bar{u}u\rangle, \  \sigma_d=\langle\bar{d}d\rangle, \  \sigma_s=\langle\bar{s}s\rangle.
	\end{eqnarray}
	
The thermodynamic potential in mean field level contains the mean field part and quark part
	\begin{eqnarray}
		\label{omega1}
		\Omega &=&\sum_{f=u,d,s}2 G\sigma_f^2-4K\sigma_u\sigma_d\sigma_s+{\cal U}(\Phi,\bar\Phi)+\Omega_q, \\
		\Omega_q &=& - \sum_{f=u,d,s}\frac{|Q_f B|}{2\pi}\sum_{l} \left(2-\delta_{l0}\right) \int \frac{d p_z}{2\pi} \Bigg[3E_f \nonumber\\
		&+& T\ln\left(1+3\Phi e^{-\beta E_f^+}+3{\bar \Phi}e^{-2\beta E_f^+}+e^{-3\beta E_f^+}\right)\nonumber\\
&+& T\ln\left(1+3{\bar \Phi} e^{-\beta E_f^-}+3{ \Phi}e^{-2\beta E_f^-}+e^{-3\beta E_f^-}\right)\Bigg],\nonumber
	\end{eqnarray}
where $E_f^\pm=E_f \pm \mu_f$ contains quark energy $E_f=\sqrt{p^2_z+2 l |Q_f B|+m_f^2}$ of longitudinal momentum $p_z$, Landau level $l$, and effective quark masses %$f=u,d,s$ means quark flavors, $l$ Landau levels, $\alpha_l=2-\delta_{l0}$ spin factor and
\begin{eqnarray}
m_u&=m_0^u-4G\sigma_u+2K\sigma_d\sigma_s,\nonumber \\
m_d&=m_0^d-4G\sigma_d+2K\sigma_u\sigma_s,\nonumber \\
m_s&=m_0^s-4G\sigma_s+2K\sigma_u\sigma_d,\nonumber
\end{eqnarray}
and quark chemical potential
\begin{eqnarray}
\mu_u&=&\frac{\mu_B}{3}+\frac{2\mu_Q}{3},\nonumber \\
\mu_d&=&\frac{\mu_B}{3}-\frac{\mu_Q}{3},\nonumber \\
\mu_s&=&\frac{\mu_B}{3}-\frac{\mu_Q}{3}-\mu_S,\nonumber
\end{eqnarray}
with $\mu_B,\ \mu_Q,\ \mu_S$ the chemical potential corresponding to the baryon number $B$, electric charge $Q$ and strangeness $S$, respectively.

The ground state at finite temperature, chemical potential and magnetic field is determined by minimizing the thermodynamic potential
\begin{eqnarray}
 \frac{\partial \Omega }{\partial \sigma_f}&=&0,\ (f=u,d,s),\nonumber\\
 \frac{\partial \Omega }{\partial \Phi}&=&0,\label{gapeqs}\\
  \frac{\partial \Omega }{\partial {\bar \Phi}}&=&0.\nonumber
\end{eqnarray}
The thermodynamic potential $\Omega$ is a function of order parameters (chiral condensates $\sigma_f$ and Polyakov loop $\Phi, {\bar \Phi}$), and hence we obtain five coupled gap equations.

At vanishing quark chemical potential $(\mu_B=\mu_Q=\mu_S=0)$ and finite temperature, the chiral symmetry restoration and deconfinement process are smooth crossover. By considering the derivative of the order parameters (chiral condensates and Polyakov loop) with respect to the temperature, the pseudo-critical temperatures $T_{pc}^c$ and $T_{pc}^s$ of chiral restoration phase transitions of light ($u/d$) quarks and strange quarks, and the pseudo-critical temperature $T_{pc}^d$ of deconfinement phase transition are determined by the location of the peak of $\frac{d \sigma_{ud}}{d T}\ (\sigma_{ud}=\frac{\sigma_u+\sigma_d}{2})$, $\frac{d \sigma_{s}}{d T}$ and $\frac{d \Phi}{d T}$ $(\Phi={\bar {\Phi}})$, respectively. Due to the heavier current mass of strange quarks, the chiral restoration phase transition usually refers to the light quarks. The strength of chiral restoration and deconfinement phase transitions is characterized by the peak value of $\frac{d \sigma_{ud}}{d T}$ and $\frac{d \Phi}{d T}$, respectively.

The correlations and fluctuations of baryon number $B$, electric charge $Q$ and strangeness $S$ can be obtained by taking the derivatives of the thermodynamic potential $\Omega$ with respect to the chemical potentials $\hat{\mu}_X=\mu_X/T,\ (X=B,\ Q,\ S)$, evaluated at zero quark chemical potential
\begin{eqnarray}
\chi_{i,j,k}^{B,Q,S}&=&-\frac{\partial^{i+j+k}(\Omega/T^4)}{\partial \hat{\mu}_B^i \partial \hat{\mu}_Q^j \partial \hat{\mu}_S^k}{\bigg |}_{{\mu}_X=0}.
\end{eqnarray}
In this work, we focus on the fourth order correlations $\chi^{BQ}_{31}$, $\chi^{QB}_{31}$, $\chi^{BQ}_{22}$, $\chi^{BS}_{31}$, $\chi^{SB}_{31}$, $\chi^{BS}_{22}$, $\chi^{QS}_{31}$, $\chi^{SQ}_{31}$, $\chi^{QS}_{22}$, $\chi^{BQS}_{211}$, $\chi^{QBS}_{211}$, $\chi^{SBQ}_{211}$ at finite temperature, magnetic field and vanishing quark chemical potential. To understand the property of correlations along the phase transition line in $T-eB$ plane, the scaled correlations ${\hat {\chi}}_{mn}^{XY}=\frac{\chi_{mn}^{XY}({\text {eB}},T_{pc}^c({\text {eB}}))}{\chi_{mn}^{XY}({\text {eB}}=0,T_{pc}^c({\text {eB}}=0))}$ ($n+m=4,\ (n,m>0)$) and ${\hat {\chi}}_{211}^{XYZ}=\frac{\chi_{211}^{XYZ}({\text {eB}},T_{pc}^c({\text {eB}}))}{\chi_{211}^{XYZ}({\text {eB}}=0,T_{pc}^c({\text {eB}}=0))}$, with $X,\ Y,\ Z=B,\ Q,\ S,\ (X \neq Y \neq Z)$ at the pseudo-critical temperature $T_{pc}^c$ of chiral restoration phase transition are also discussed, which is analogous to studying the central-to-peripheral collision conditions. Note that we have similar pseudo-critical temperature for chiral restoration and deconfinement phase transitions $T_{pc}^c \simeq T_{pc}^d$ (shown in Table.\ref{tabletpc}). Together with our previous work~\cite{maoyangprd}, we shows a comprehensive results of second and fourth order correlations and fluctuations of conserved charges at finite magnetic field and temperature.

\section{Numerical Results}
\label{results}

\subsection{fourth order correlations}
Because of the contact interaction in NJL model, the ultraviolet divergence cannot be eliminated through renormalization, and we apply the covariant Pauli-Villars regularization~\cite{mao}. In this scheme, the quark momentum and Landau level run formally from zero to infinity, and the divergence is removed by the cancellation among the subtraction terms. One introduces the regularized quark masses $m_{fi}=\sqrt{m^2_f+a_i\Lambda^2}$ for $i=0,1,\cdots, N$, and replaces $m^2_f$ in the quark energy $E_f=\sqrt{p^2_z+2l|Q_fB|+m^2_f}$ by $m_{fi}^2$. And the summation and integration in thermodynamical potential in Eq.\eqref{omega1} and the corresponding gap equations in Eq.\eqref{gapeqs} are changed as
\begin{eqnarray}
&&\sum_l\int \frac{dp_z}{2\pi} Function(E_f)\ \ \ \longrightarrow \nonumber\\
&&\sum_l\int \frac{dp_z}{2\pi} \sum_{i=0}^N \left[ c_i \times Function(E_f^i) \right], \nonumber
\end{eqnarray}
with $E_f^i=\sqrt{p^2_z+2l|Q_fB|+m^2_{fi}}$. The parameters $N=3$, $a_1=1, c_1=-3$, $a_2=2, c_2=3$, $a_3=3, c_3=-1$, are determined by constraints $a_0=0$, $c_0=1$, and $\sum_{i=0}^N c_im_{fi}^{2L}=0$ for $L=0,1,\cdots N-1$.

By fitting the physical quantities, pion mass $m_{\pi}=138\ \text{MeV}$, pion decay constant $f_{\pi}=93\ \text{MeV}$, kaon mass $m_K=495.7\ \text{MeV}$, $\eta'$ meson mass $m_{\eta\prime}=957.5\ \text{MeV}$ in vacuum, we fix the current masses of light quarks $m_0^{u}=m_0^{d}=5.5\ \text{MeV}$, and obtain the parameters $m_0^s=154.7\ \text{MeV}$, $G\Lambda^2=3.627$, $K\Lambda^5=92.835$, $\Lambda=1101\ \text{MeV}$~\cite{3pnjlmei,maoyangprd}. For the Polyakov potential, the parameters are chosen as~\cite{pnjl6,maoyangprd} $b_{20}=6.75$, $b_{21}=-1.95$, $b_{22}=2.625$, $b_{23}=-7.44$, $b_3=0.75$, $b_4=7.5$, and $T_0=270$ MeV. The resulting pseudo-critical temperatures of chiral restoration and deconfinement phase transitions are listed in Table.\ref{tabletpc}, with $T^c_{pc} \simeq T^d_{pc} < T^s_{pc}$ at vanishing and non-vanishing magnetic field. The strength of chiral restoration and deconfinement phase transitions increase with increasing magnetic field, as shown in Fig.\ref{figebpara}.% is consistent with LQCD results~\cite{lattice1,lattice2,lattice2sep}

%%%%%%%%%%%%%%%%%%%%%%%%%%%%%%%%%
\begin{table}
\begin{tabular}{cccc}
\hline
\hline
                      & $eB=0$ \               & $eB=10 (m_{\pi}^2)$\         & $eB=20 (m_{\pi}^2)$  \\
\hline
$T^c_{pc} ({\text {MeV}})$  & 205  \               &   219    \         & 225\\
\hline
$T^d_{pc} ({\text {MeV}})$  &  210  \              &    215 \            &219\\
\hline
$T^s_{pc} ({\text {MeV}})$  &  260  \              &   262  \          & 264\\
\hline

\hline
\end{tabular}
\caption{Pseudo-critical temperatures of chiral restoration and deconfinement phase transitions in three-flavor magnetized PNJL model with the original parameters.}
\label{tabletpc}
\end{table}
%%%%%%%%%%%%%%%%%%%%%%%%%%%%%

%%%%%%%%%%%%%%%%%%%%%%%%%%%%%%%%%%%%%%%%%%%%%%%%%%%%%%%%%%%%%%%%%%%%%%%%%%%%%%%%%%%%%%%%%%%%%%%%%%%%%%%%%%%%%%%%%%%%%%%%%%%%%%%%%%%%%%%%%%%%%%%%%%%%
\begin{figure}[htb]
\begin{center}
\includegraphics[width=6cm]{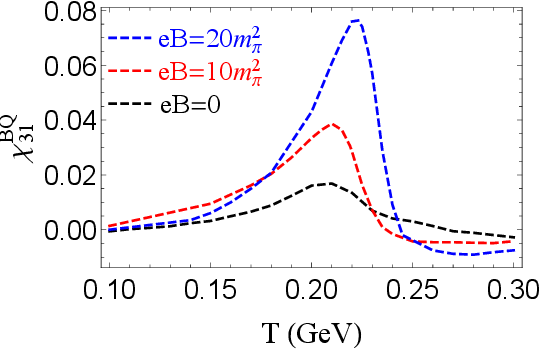}
\includegraphics[width=6cm]{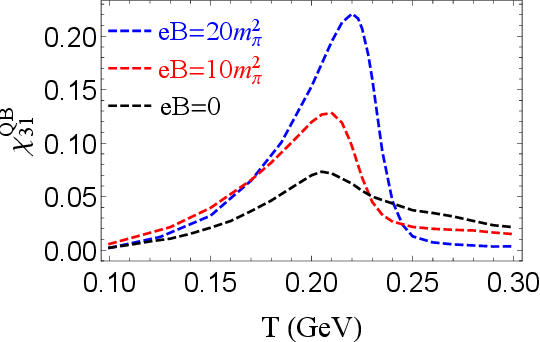}
\includegraphics[width=6cm]{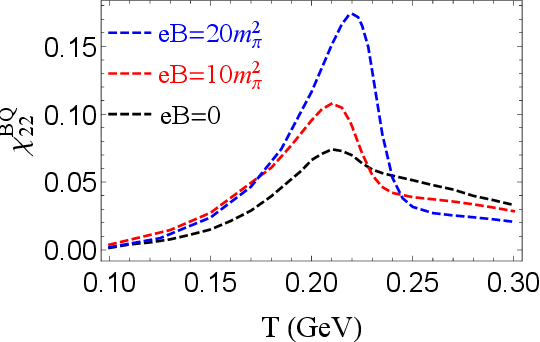}
\includegraphics[width=5.75cm]{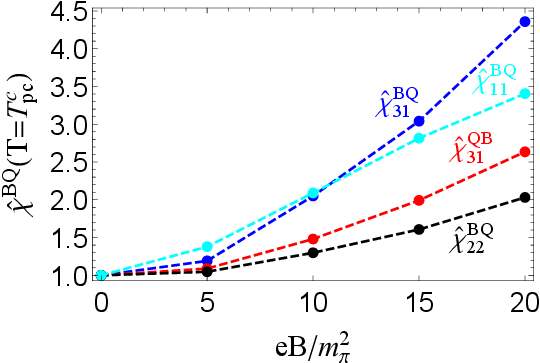}
\end{center}
\caption{The fourth order correlations $\chi^{BQ}_{31}$ (upper panel), $\chi^{QB}_{31}$ (second panel) and $\chi^{BQ}_{22}$ (third panel) of baryon number $B$ and electric charge $Q$ as functions of temperature with and without magnetic field. (bottom panel) The scaled fourth and second order correlations $\hat{\chi}^{BQ}=\frac{\chi^{BQ}({\text {eB}},T_{pc}^c({\text {eB}}))}{\chi^{BQ}({\text {eB}}=0,T_{pc}^c({\text {eB}}=0))}$ of baryon number and electric charge at the pseudo-critical temperature $T^c_{pc}$ of light quark chiral restoration as functions of magnetic field.}
\label{figxbq}
\end{figure}
%%%%%%%%%%%%%%%%%%%%%%%%%%%%%%%%%%%%%%%%%%%%%%%%%%%%%%%%%%%%%%%%%%%%%%%

%%%%%%%%%%%%%%%%%%%%%%%%%%%%%%%%%%%%%%%%%%%%%%%%%%%%%%%%%%%%%%%%%%%%%%%%%%%%%%%%%%%%%%%%%%%%%%%%%%%%%%%%%%%%%%%%%%%%%%%%%%%%%%%%%%%%%%%%%%%%%%%%%%%%
\begin{figure}[htb]
\begin{center}
\includegraphics[width=6cm]{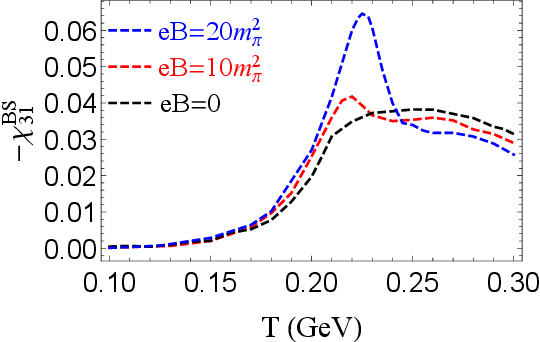}
\includegraphics[width=6cm]{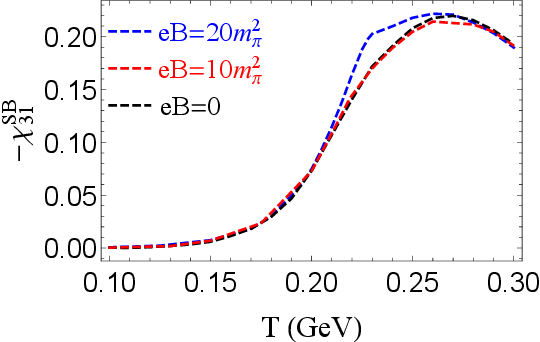}
\includegraphics[width=6cm]{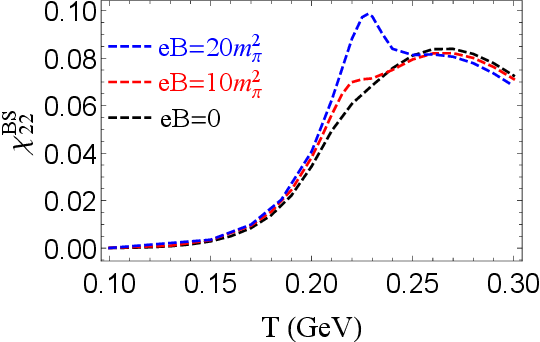}
\includegraphics[width=5.75cm]{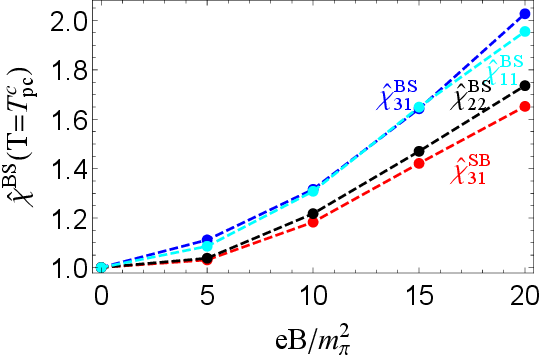}
\end{center}
\caption{The fourth order correlations $-\chi^{BS}_{31}$ (upper panel), $-\chi^{SB}_{31}$ (second panel) and $\chi^{BS}_{22}$ (third panel) of baryon number $B$ and strangeness $S$ as functions of temperature with and without magnetic field. (bottom panel) The scaled fourth and second order correlations $\hat{\chi}^{BS}=\frac{\chi^{BS}({\text {eB}},T_{pc}^c({\text {eB}}))}{\chi^{BS}({\text {eB}}=0,T_{pc}^c({\text {eB}}=0))}$ of baryon number and strangeness  at the pseudo-critical temperature $T^c_{pc}$ of light quark chiral restoration as functions of magnetic field.}
\label{figxbs}
\end{figure}
%%%%%%%%%%%%%%%%%%%%%%%%%%%%%%%%%%%%%%%%%%%%%%%%%%%%%%%%%%%%%%%%%%%%%%%
%%%%%%%%%%%%%%%%%%%%%%%%%%%%%%%%%%%%%%%%%%%%%%%%%%%%%%%%%%%%%%%%%%%%%%%%%%%%%%%%%%%%%%%%%%%%%%%%%%%%%%%%%%%%%%%%%%%%%%%%%%%%%%%%%%%%%%%%%%%%%%%%%%%%
\begin{figure}[htb]
\begin{center}
\includegraphics[width=6cm]{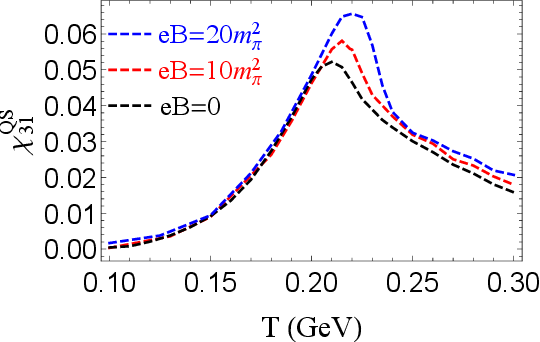}
\includegraphics[width=6cm]{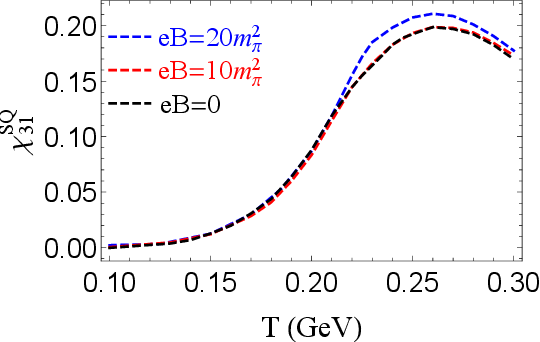}
\includegraphics[width=6cm]{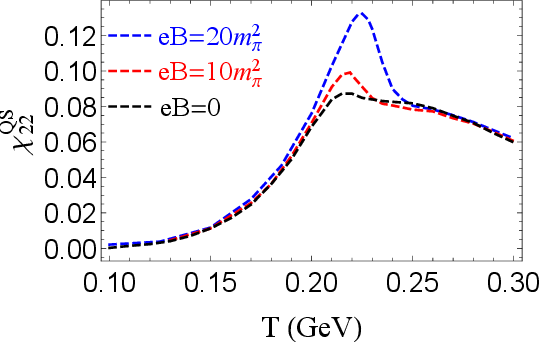}
\includegraphics[width=5.75cm]{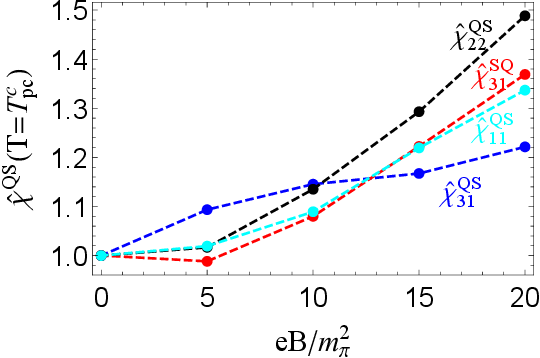}
\end{center}
\caption{The fourth order correlations $\chi^{QS}_{31}$ (upper panel), $\chi^{SQ}_{31}$ (second panel) and $\chi^{QS}_{22}$ (third panel) of electric charge $Q$ and strangeness $S$ as functions of temperature with and without magnetic field. (bottom panel) The scaled fourth and second order correlations $\hat{\chi}^{QS}=\frac{\chi^{QS}({\text {eB}},T_{pc}^c({\text {eB}}))}{\chi^{QS}({\text {eB}}=0,T_{pc}^c({\text {eB}}=0))}$ of electric charge and strangeness at the pseudo-critical temperature $T^c_{pc}$ of light quark chiral restoration as functions of magnetic field.}
\label{figxqs}
\end{figure}
%%%%%%%%%%%%%%%%%%%%%%%%%%%%%%%%%%%%%%%%%%%%%%%%%%%%%%%%%%%%%%%%%%%%%%%
%%%%%%%%%%%%%%%%%%%%%%%%%%%%%%%%%%%%%%%%%%%%%%%%%%%%%%%%%%%%%%%%%%%%%%%%%%%%%%%%%%%%%%%%%%%%%%%%%%%%%%%%%%%%%%%%%%%%%%%%%%%%%%%%%%%%%%%%%%%%%%%%%%%%
\begin{figure}[htb]
\begin{center}
\includegraphics[width=6cm]{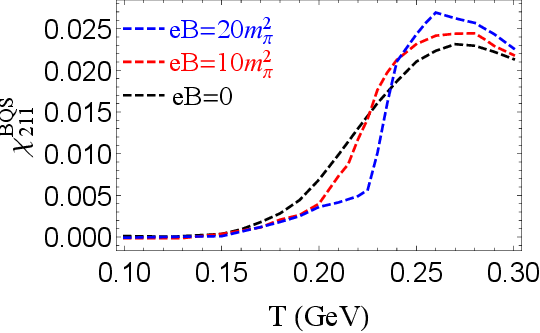}
\includegraphics[width=6cm]{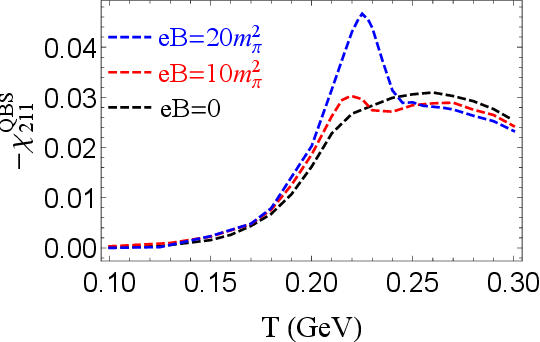}
\includegraphics[width=6cm]{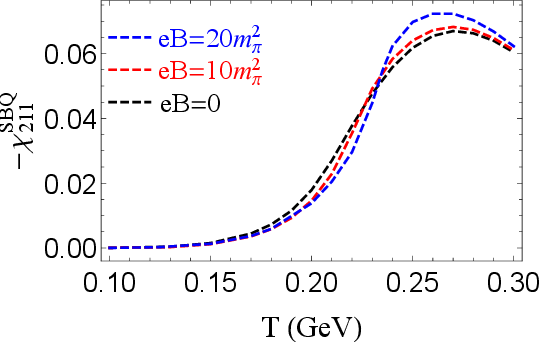}
\includegraphics[width=5.75cm]{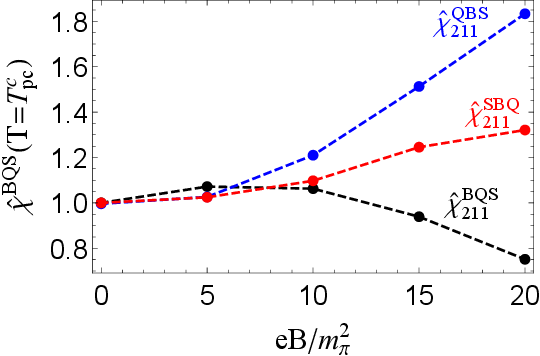}
\end{center}
\caption{The fourth order correlations $\chi^{BQS}_{211}$ (upper panel), $-\chi^{QBS}_{211}$ (second panel) and $-\chi^{SBQ}_{211}$ (third panel) among baryon number $B$, electric charge $Q$ and strangeness $S$ as functions of temperature with and without magnetic field. (bottom panel) The scaled fourth order correlations $\hat{\chi}^{BQS}=\frac{\chi^{BQS}({\text {eB}},T_{pc}^c({\text {eB}}))}{\chi^{BQS}({\text {eB}}=0,T_{pc}^c({\text {eB}}=0))}$ among baryon number, electric charge and strangeness at the pseudo-critical temperature $T^c_{pc}$ of light quark chiral restoration as functions of magnetic field.}
\label{figxbqs}
\end{figure}
%%%%%%%%%%%%%%%%%%%%%%%%%%%%%%%%%%%%%%%%%%%%%%%%%%%%%%%%%%%%%%%%%%%%%%%

Figure \ref{figxbq} shows the results of the fourth order correlations $\chi^{BQ}_{31}$, $\chi^{QB}_{31}$ and $\chi^{BQ}_{22}$ of baryon number $B$ and electric charge $Q$ at finite temperature and magnetic field. With vanishing and non-vanishing magnetic field, the fourth order correlations $\chi^{BQ}_{31}$, $\chi^{QB}_{31}$ and $\chi^{BQ}_{22}$ increases with temperature, and have a peak around the pseudo-critical temperature $T^c_{pc}$ of light quark chiral restoration. With stronger magnetic field, there exists a sharper peak. The scaled fourth order correlations ${\hat {\chi}}_{mn}^{BQ}\ (n+m=4)$ and scaled second order correlation ${\hat {\chi}}_{11}^{BQ}$ at the pseudo-critical temperature $T^c_{pc}$ of light quark chiral restoration are plotted in the bottom panel of Fig.\ref{figxbq}. They increase with magnetic field, which is caused by the increasing phase transition strength under external magnetic field. Among them, ${\hat {\chi}}_{11}^{BQ}$ increases fastest at weak magnetic field $eB<10 m_{\pi}^2$ and ${\hat {\chi}}_{31}^{BQ}$ increases fastest at strong magnetic field $eB>10 m_{\pi}^2$.

Figure \ref{figxbs} plots the results of the fourth order correlations $-\chi^{BS}_{31}$, $-\chi^{SB}_{31}$ and $\chi^{BS}_{22}$ of baryon number $B$ and strangeness $S$ at finite temperature and magnetic field. With vanishing magnetic field, fourth order correlations $-\chi^{BS}_{31}$, $-\chi^{SB}_{31}$ and $\chi^{BS}_{22}$ increase with temperature and show wide peak around the pseudo-critical temperature $T^s_{pc}$ of strange quark chiral restoration. Since the strangeness is contributed only by strange quarks. When turning on magnetic fields, another peak appears around the pseudo-critical temperature $T^c_{pc}$ of light quark chiral restoration. The peak becomes more pronounced with the increase of magnetic field. The scaled fourth order correlations ${\hat {\chi}}_{mn}^{BS}\ (n+m=4)$ and scaled second order correlation ${\hat {\chi}}_{11}^{BS}$ at the pseudo-critical temperature $T^c_{pc}$ of light quark chiral restoration are plotted in the bottom panel of Fig.\ref{figxbs}. They increases with magnetic field, which is caused by the increasing phase transition strength under external magnetic field. Among them, ${\hat {\chi}}_{31}^{BS}$ and ${\hat {\chi}}_{11}^{BS}$ increase fastest at finite magnetic field. Comparing with the scaled correlations ${\hat {\chi}}_{31}^{BQ}$ and ${\hat {\chi}}_{11}^{BQ}$ in Fig.\ref{figxbq}, ${\hat {\chi}}_{31}^{BS}$ and ${\hat {\chi}}_{11}^{BS}$ are less sensitive to the magnetic field. The reason lies in the heavier mass of strange quarks than light ($u/d$) quarks.

Figure \ref{figxqs} depicts the results of the fourth order correlations $\chi^{QS}_{31}$, $\chi^{SQ}_{31}$ and $\chi^{QS}_{22}$ of electric charge $Q$ and strangeness $S$ at finite temperature and magnetic field. With vanishing and non-vanishing magnetic field, $\chi^{QS}_{31}$ and $\chi^{QS}_{22}$ increase with temperature and have the peak structure around the pseudo-critical temperature $T^c_{pc}$ of light quark chiral restoration. The peak becomes sharper with increasing magnetic field. With vanishing and non-vanishing magnetic field, $\chi^{SQ}_{31}$ shows wide peak around the pseudo-critical temperature $T^s_{pc}$ of strange quark chiral restoration, which is slightly enhanced by magnetic field. The scaled fourth order correlations ${\hat {\chi}}_{mn}^{QS}\ (n+m=4)$ and scaled second order correlation ${\hat {\chi}}_{11}^{QS}$ at the pseudo-critical temperature $T^c_{pc}$ of light quark chiral restoration are plotted in the bottom panel of Fig.\ref{figxqs}. $\chi^{QS}_{31}$, $\chi^{QS}_{22}$ and $\chi^{QS}_{11}$ go up as the magnetic field grows. $\chi^{SQ}_{31}$ firstly decreases and then increases with magnetic field. Among them, ${\hat {\chi}}_{22}^{QS}$ ($\chi^{QS}_{31}$) increases fastest in strong magnetic field case $eB>10m_\pi^2$ (in weak magnetic field case $eB<10m_\pi^2$). But they are less sensitive to the magnetic field than ${\hat {\chi}}_{31}^{BQ}$, ${\hat {\chi}}_{11}^{BQ}$ in Fig.\ref{figxbq} and ${\hat {\chi}}_{31}^{BS}$, ${\hat {\chi}}_{11}^{BS}$ in Fig.\ref{figxbs}.

Figure \ref{figxbqs} shows the fourth order correlations $\chi^{BQS}_{211}$, $-\chi^{QBS}_{211}$ and $-\chi^{SBQ}_{211}$ among baryon number $B$, electric charge $Q$ and strangeness $S$ at finite temperature and magnetic field. With vanishing and non-vanishing magnetic field, $\chi^{BQS}_{211}$ and $-\chi^{SBQ}_{211}$ increase with temperature, and the peak structure appears around the pseudo-critical temperature $T^s_{pc}$ of strange quark chiral restoration. $-\chi^{QBS}_{211}$ has a wide peak around the pseudo-critical temperature $T^s_{pc}$ of strange quark chiral restoration at vanishing magnetic field. Another peak shows up around the pseudo-critical temperature $T^c_{pc}$ of light quark chiral restoration at finite magnetic field, and at strong enough magnetic field, there only exists a peak around the pseudo-critical temperature $T^c_{pc}$ of light quark chiral restoration. The scaled fourth order correlations $\hat{\chi}^{BQS}$ at the pseudo-critical temperature $T^c_{pc}$ of light quark chiral restoration is plotted as a function of magnetic field in the bottom panel of Fig.\ref{figxbqs}. $\hat{\chi}^{QBS}_{211}$ increases fastest with magnetic field, but $\hat{\chi}^{BQS}_{211}$ decreases with magnetic field.

\begin{figure*}[htb]
\begin{center}
\includegraphics[width=6.3cm]{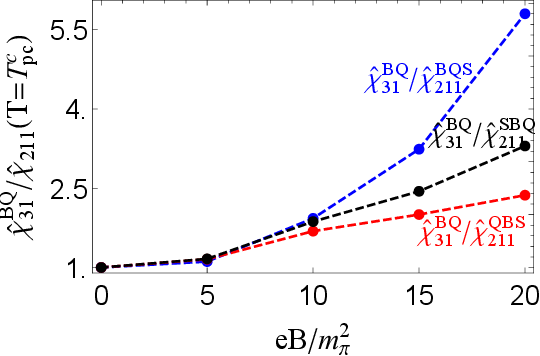}\includegraphics[width=6.3cm]{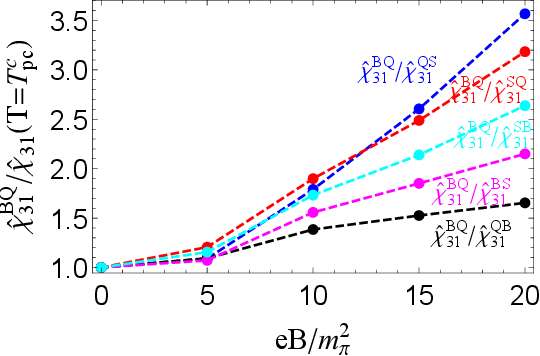}\\
\includegraphics[width=6.3cm]{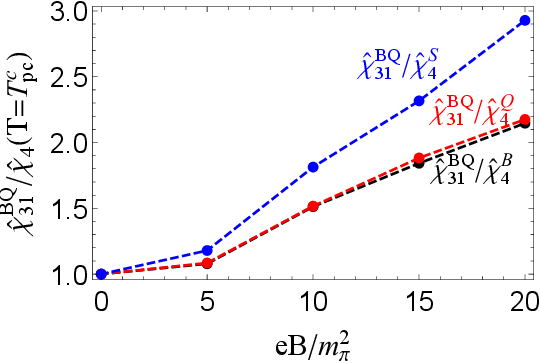}\includegraphics[width=6.3cm]{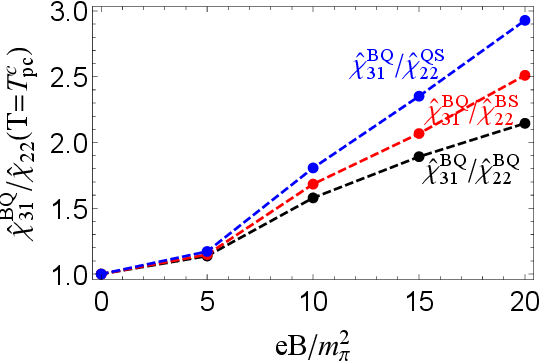}\\
\includegraphics[width=6.3cm]{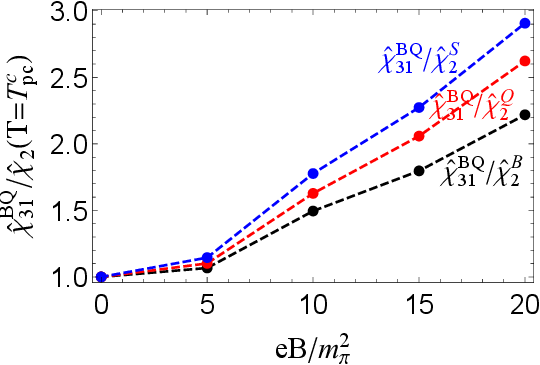}\includegraphics[width=6.3cm]{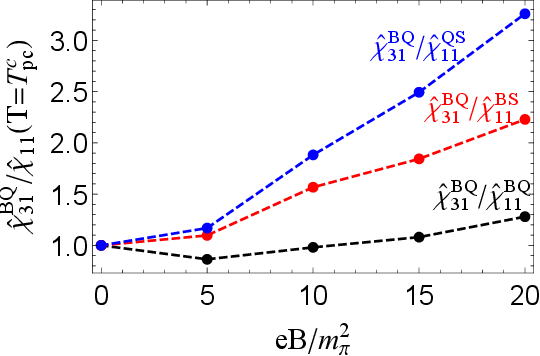}
\end{center}
\caption{Double ratio ${\cal R}={\hat {\chi}}_{(2)}/{\hat {\chi}}_{(1)}$ as functions of magnetic field. Here, ${\hat {\chi}}_{31}^{BQ}$ is selected as ${\hat {\chi}}_{(2)}$, and other correlations and fluctuations as ${\hat {\chi}}_{(1)}$.}
\label{figx4x4tpc}
\end{figure*}

Based on the above results, we plot the double ratio ${\cal R}={\hat {\chi}}_{(2)}/{\hat {\chi}}_{(1)}$ as functions of magnetic field in Fig.\ref{figx4x4tpc}. Such ratios are widely adopted in heavy-ion experiments to suppress volume effect and analyzed as functions of centrality~\cite{ding99,ding100}. Given the fact that magnetic field strength grows from central to peripheral collisions, ${\cal R}$ can be helpful to probe the magnetic field. To amplify magnetic field signature, we select ${\hat {\chi}}_{31}^{BQ}$ as ${\hat {\chi}}_{(2)}$, which is most sensitive to the magnetic field. The figures of ${\cal R}$ are arranged by their sensitivity to magnetic field when choosing other correlations and fluctuations as ${\hat {\chi}}_{(1)}$. ${\cal R}$ in Fig.\ref{figx4x4tpc} increases with magnetic field, except ${\hat {\chi}}_{31}^{BQ}/{\hat {\chi}}_{11}^{BQ}$. Among them, ${\hat {\chi}}_{31}^{BQ}/{\hat {\chi}}_{211}^{BQS}$ increases fastest. ${\hat {\chi}}_{31}^{BQ}/{\hat {\chi}}_{11}^{BQ}$ decreases in weak magnetic field and then increases in strong magnetic field.

%%%%%%%%%%%%%%%%%%%%%%%%%%%%%%%%%%%%%%%%%%%%%%%%%%%%%%%%%%%%%%%%%%%%%%%%%%%%%%%%%%%%%%%%%%%%%%%%%%%%%%%%%%%%%%%%%%%%%%%%%%%%%%%%%%%%%%%%%%%%%%%%%%%%
%%%%%%%%%%%%%%%%%%%%%%%%%%%%%%%%%%%%%%%%%%%%%%%%%%%%%%%%%%%%%%%%%%%%%%%

\subsection{IMC effect}

%%%%%%%%%%%%%%%%%%%%%%%%%%%%%%%%%%%%%%%%%%%%%%%%%%%%%%%%%%%%%%%%%%%%%%%%%%%%%%%%%%%%%%%%%%%%%%%%%%%%%%%%%%%%%%%%%%%%%%%%%%%%%%%%%%%%%%%%%%%%%%%%%%%%
\begin{figure}[htb]
\begin{center}
\includegraphics[width=6.4cm]{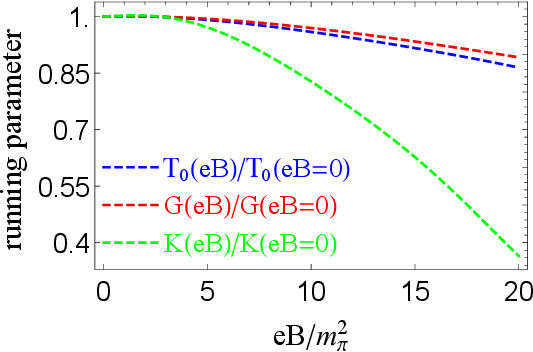}
\includegraphics[width=6.5cm]{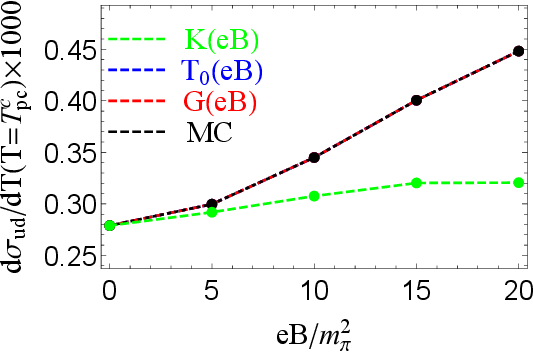}
\end{center}
\caption{(upper panel) Magnetic field dependent parameters $G(eB)$ (red line), $K(eB)$ (green line) and $T_0(eB)$ (blue line), fitted from LQCD reported decreasing pseudo-critical temperature of chiral restoration phase transition $T_{pc}^c(eB)/T_{pc}^c(eB=0)$ under external magnetic field~\cite{lattice1}. (lower panel) Strength of chiral restoration phase transition ($\frac{d \sigma_{ud}}{d T}$ at $T=T_{pc}^c$) under external magnetic field with IMC effect (green, blue and red lines) and without IMC effect (black line).}
\label{figebpara}
\end{figure}
%%%%%%%%%%%%%%%%%%%%%%%%%%%%%%%%%%%%%%%%%%%%%%%%%%%%%%%%%%%%%%%%%%%%%%%
%%%%%%%%%%%%%%%%%%%%%%%%%%%%%%%%%%%%%%%%%%%%%%%%%%%%%%%%%%%%%%%%%%%%%%%%%%%%%%%%%%%%%%%%%%%%%%%%%%%%%%%%%%%%%%%%%%%%%%%%%%%%%%%%%%%%%%%%%%%%%%%%%%%%
\begin{figure}[htb]
\begin{center}
\includegraphics[width=6cm]{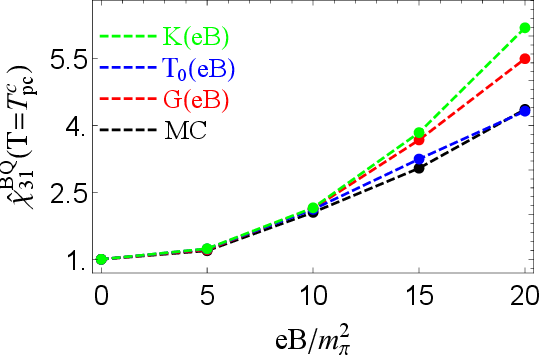}
\includegraphics[width=6cm]{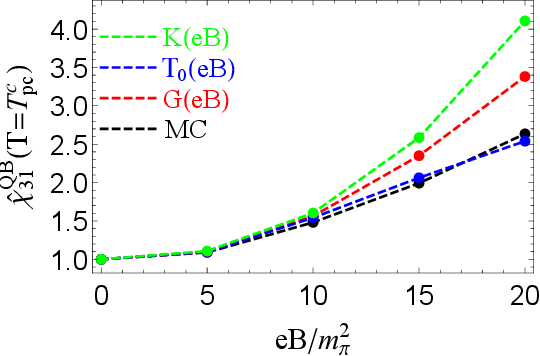}
\includegraphics[width=6cm]{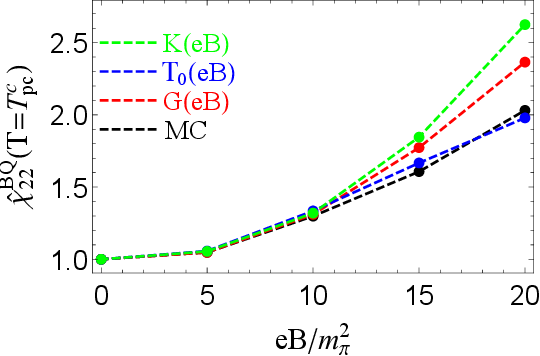}
\includegraphics[width=6cm]{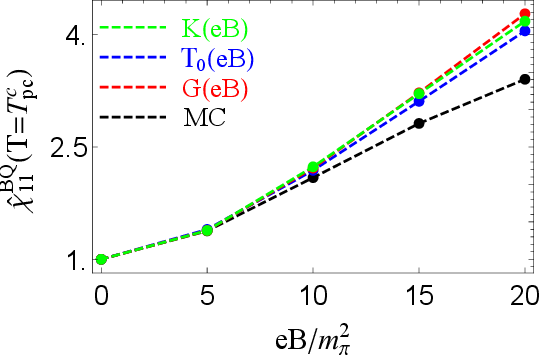}
\end{center}
\caption{The scaled fourth and second order correlations of baryon number $B$ and electric charge $Q$ at pseudo-critical temperature of chiral restoration phase transition as functions of magnetic field with and without IMC effect. For all panels, the green, blue and red lines are the results in case of considering IMC effect in $K(eB)$, $T_0(eB)$ and $G(eB)$ scheme, respectively, and the black lines are the results without IMC effect.}
\label{figxbqtpcimc}
\end{figure}
%%%%%%%%%%%%%%%%%%%%%%%%%%%%%%%%%%%%%%%%%%%%%%%%%%%%%%%%%%%%%%%%%%%%%%%
%%%%%%%%%%%%%%%%%%%%%%%%%%%%%%%%%%%%%%%%%%%%%%%%%%%%%%%%%%%%%%%%%%%%%%%%%%%%%%%%%%%%%%%%%%%%%%%%%%%%%%%%%%%%%%%%%%%%%%%%%%%%%%%%%%%%%%%%%%%%%%%%%%%%
\begin{figure}[htb]
\begin{center}
\includegraphics[width=6cm]{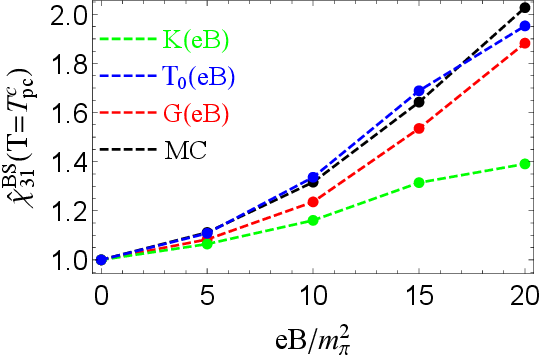}
\includegraphics[width=6cm]{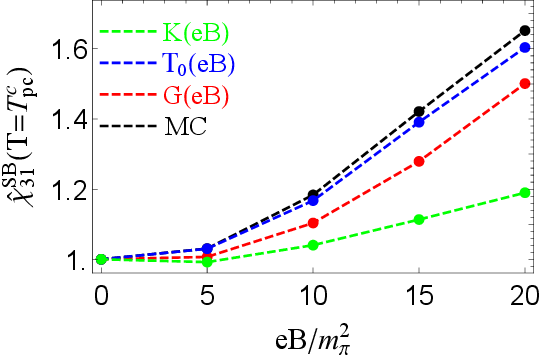}
\includegraphics[width=6cm]{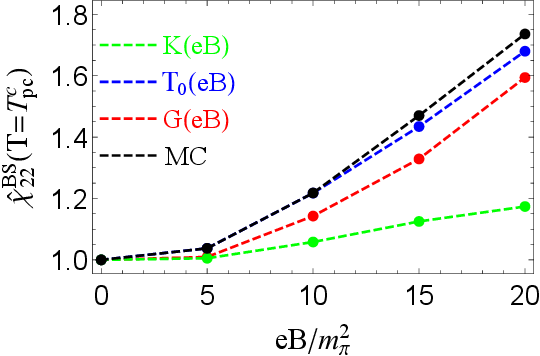}
\includegraphics[width=6cm]{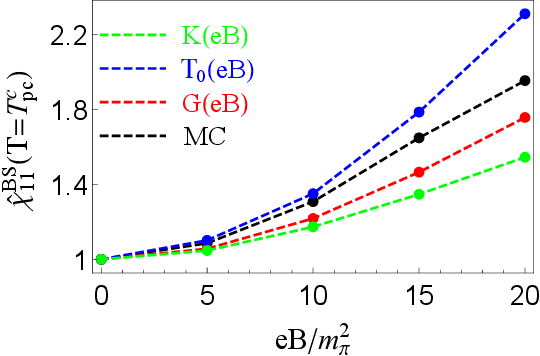}
\end{center}
\caption{The scaled fourth and second order correlations of baryon number $B$ and strangeness $S$ at pseudo-critical temperature of chiral restoration phase transition as functions of magnetic field with and without IMC effect. For all panels, the green, blue and red lines are the results in case of considering IMC effect in $K(eB)$, $T_0(eB)$ and $G(eB)$ scheme, respectively, and the black lines are the results without IMC effect.}
\label{figxbstpcimc}
\end{figure}
%%%%%%%%%%%%%%%%%%%%%%%%%%%%%%%%%%%%%%%%%%%%%%%%%%%%%%%%%%%%%%%%%%%%%%%
%%%%%%%%%%%%%%%%%%%%%%%%%%%%%%%%%%%%%%%%%%%%%%%%%%%%%%%%%%%%%%%%%%%%%%%%%%%%%%%%%%%%%%%%%%%%%%%%%%%%%%%%%%%%%%%%%%%%%%%%%%%%%%%%%%%%%%%%%%%%%%%%%%%%
\begin{figure}[htb]
\begin{center}
\includegraphics[width=6cm]{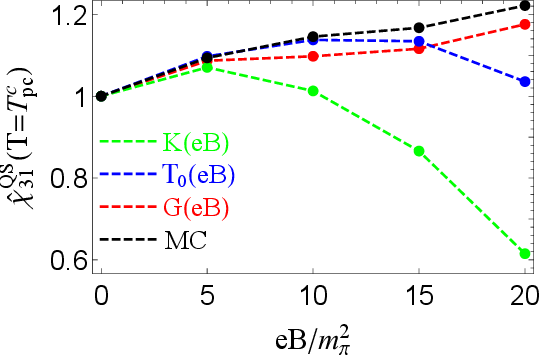}
\includegraphics[width=6cm]{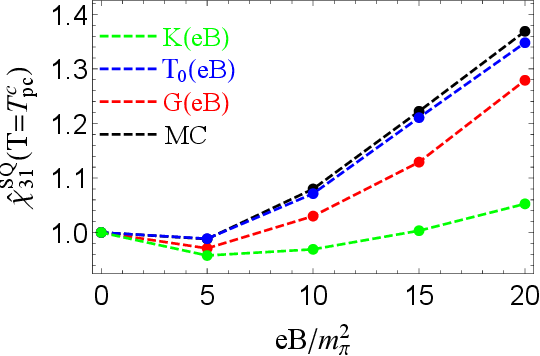}
\includegraphics[width=6cm]{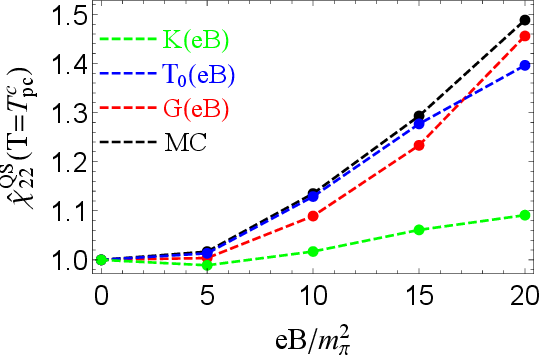}
\includegraphics[width=6cm]{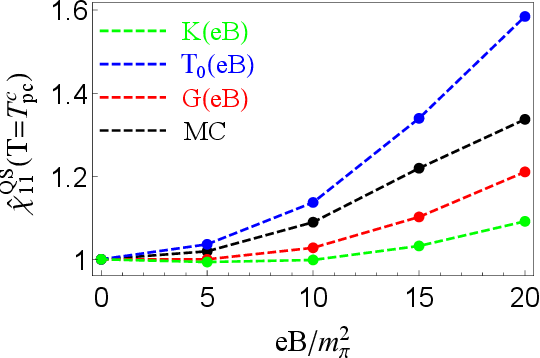}
\end{center}
\caption{The scaled fourth and second order correlations of electric charge $Q$ and strangeness $S$ at pseudo-critical temperature of chiral restoration phase transition as functions of magnetic field with and without IMC effect. For all panels, the green, blue and red lines are the results in case of considering IMC effect in $K(eB)$, $T_0(eB)$ and $G(eB)$ scheme, respectively, and the black lines are the results without IMC effect.}
\label{figxqstpcimc}
\end{figure}
%%%%%%%%%%%%%%%%%%%%%%%%%%%%%%%%%%%%%%%%%%%%%%%%%%%%%%%%%%%%%%%%%%%%%%%
%%%%%%%%%%%%%%%%%%%%%%%%%%%%%%%%%%%%%%%%%%%%%%%%%%%%%%%%%%%%%%%%%%%%%%%%%%%%%%%%%%%%%%%%%%%%%%%%%%%%%%%%%%%%%%%%%%%%%%%%%%%%%%%%%%%%%%%%%%%%%%%%%%%%
\begin{figure}[htb]
\begin{center}
\includegraphics[width=6cm]{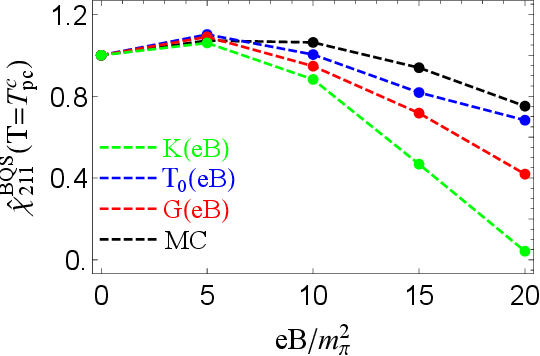}
\includegraphics[width=6cm]{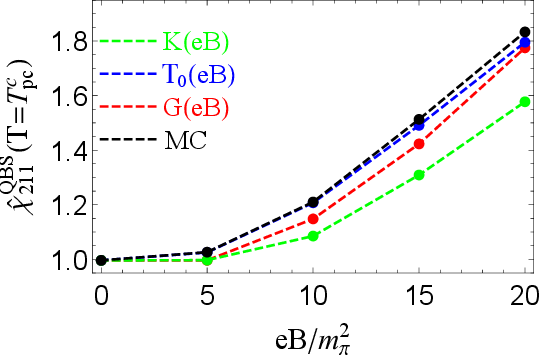}
\includegraphics[width=6cm]{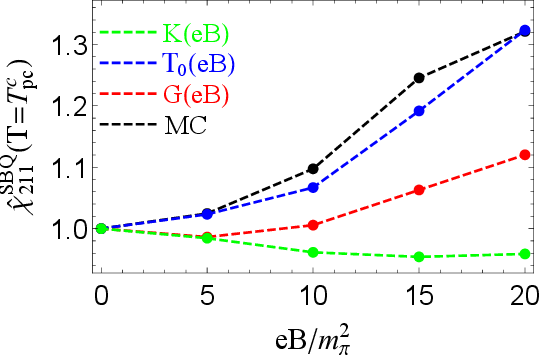}
\end{center}
\caption{The scaled fourth order correlations of baryon number $B$, electric charge $Q$ and strangeness $S$ at pseudo-critical temperature of chiral restoration phase transition as functions of magnetic field with and without IMC effect. For all panels, the green, blue and red lines are the results in case of considering IMC effect in $K(eB)$, $T_0(eB)$ and $G(eB)$ scheme, respectively, and the black lines are the results without IMC effect.}
\label{figxbqstpcimc}
\end{figure}
%%%%%%%%%%%%%%%%%%%%%%%%%%%%%%%%%%%%%%%%%%%%%%%%%%%%%%%%%%%%%%%%%%%%%%%

The typical results of PNJL model with the original parameters determined from the vacuum properties present the magnetic catalysis phenomena at finite temperature and vanishing quark chemical potential. The order parameters, chiral condensates $\sigma_{f=u,d,s}$ and Polyakov loop $\Phi, {\bar \Phi}$, increase with the magnetic field in the whole temperature region. The pseudo-critical temperatures for chiral restoration and deconfinement phase transitions increase with the magnetic field. It should be mentioned that the catalysis phenomena of order parameters at high temperature and pseudo-critical temperature of chiral restoration and deconfinement phase transitions under external magnetic field are contrary to LQCD results~\cite{lattice1,lattice2,lattice2sep,lattice4,lattice5,lattice6,lattice7,lattice9,ding2025}.

By fitting the LQCD reported decreasing pseudo-critical temperature of chiral symmetry restoration $T_{pc}^c(eB)/T_{pc}^c(eB=0)$ under external magnetic field~\cite{lattice1}, we introduce the IMC effect in our three-flavor PNJL model through the magnetic field dependent parameter $G(eB)$, $K(eB)$ and $T_0(eB)$, respectively, which represent the influence of external magnetic field to the quark-gluon interaction. On one side, the coupling between quarks plays a significant role in determining the spontaneous breaking and restoration of chiral symmetry. Considering the direct interaction between quarks and external magnetic field, a magnetic field dependent coupling $G(eB)$~\cite{bf8,bf9,maoyangprd,mao2pnjl,geb1,su3meson4,mao11,tian} or $K(eB)$~\cite{tian} is introduced into the PNJL model. On the other side, the interaction between the Polyakov loop and sea quarks may be important for the mechanism of IMC~\cite{lattice9}. A magnetic field dependent Polyakov loop scale parameter $T_0(eB)$~\cite{t0effectmao,t0effect,pnjl3,maoyangprd,mao2pnjl} is introduced into the PNJL model to mimic the reaction of the gluon sector to the presence of magnetic fields. As plotted in Fig.\ref{figebpara} upper panel, $G(eB)$, $K(eB)$ and $T_0(eB)$ are all monotonic decreasing functions of magnetic field, with $K(eB)$ of fastest decreasing slope.

We have checked that, with our fitted parameter $G(eB)$, $K(eB)$ or $T_0(eB)$, the increase (decrease) of chiral condensates with magnetic fields at the low (high) temperature, the increase of Polyakov loop with magnetic fields in the whole temperature region and the reduction of pseudo-critical temperature of deconfinement phase transition under magnetic fields can be realized. Moreover, the inclusion of IMC effect does not qualitatively change the strength of chiral restoration phase transition under external magnetic field, see Fig.\ref{figebpara} lower panel, where $\frac{d \sigma_{ud}}{d T}$ at the pseudo-critical temperature of chiral restoration phase transition increases with magnetic fields. The result with IMC effect in $G(eB)$ and $T_0(eB)$ schemes are almost same as in the cases without IMC effect, and the result with IMC effect in $K(eB)$ scheme shows lower values. This indicates that including IMC effect will not lead to qualitative difference in the results of correlations~\cite{mao2pnjl,maoyangprd}.

Figure~\ref{figxbqtpcimc} plots the scaled fourth and second order correlations of baryon number $B$ and electric charge $Q$ at pseudo-critical temperature of chiral restoration phase transition as functions of magnetic field with and without IMC effect, which increase with magnetic field. Including IMC effect does not lead to qualitative difference but causes quantitative modifications. The scaled fourth order correlations, ${\hat {\chi}}_{31}^{BQ}$, ${\hat {\chi}}_{31}^{QB}$ and ${\hat {\chi}}_{22}^{BQ}$ increase faster with IMC effect in $G(eB)$ and $K(eB)$ schemes than without IMC effect, and their results with IMC effect in $T_0(eB)$ scheme is very close to that without IMC effect. The scaled second order correlation ${\hat {\chi}}_{11}^{BQ}$ increases faster with IMC effect in $G(eB)$, $K(eB)$ and $T_0(eB)$ schemes than without IMC effect.

Figure~\ref{figxbstpcimc} plots the scaled fourth and second order correlations of baryon number $B$ and strangeness $S$ at pseudo-critical temperature of chiral restoration phase transition as functions of magnetic field with and without IMC effect, which increase with magnetic field. Including IMC effect does not lead to qualitative difference but causes quantitative modifications. The scaled fourth order correlations, ${\hat {\chi}}_{31}^{BS}$, ${\hat {\chi}}_{31}^{SB}$ and ${\hat {\chi}}_{22}^{BS}$ increase slower with IMC effect in $G(eB)$ and $K(eB)$ schemes than without IMC effect, and their results with IMC effect in $T_0(eB)$ scheme is similar to that without IMC effect. The scaled second order correlation ${\hat {\chi}}_{11}^{BS}$ increases faster (slower) with IMC effect in $T_0(eB)$ ($G(eB)$ and $K(eB)$) scheme(s) than without IMC effect.

Figure~\ref{figxqstpcimc} plots the scaled fourth and second order correlations of electric charge $Q$ and strangeness $S$ at pseudo-critical temperature of chiral restoration phase transition as functions of magnetic field with and without IMC effect. ${\hat {\chi}}_{31}^{QS}$ monotonically increases with magnetic field in case without IMC effect and with IMC effect in $G(eB)$ scheme. However, ${\hat {\chi}}_{31}^{QS}$ firstly increases and then decreases with magnetic field in case with IMC effect in $K(eB)$ and $T_0(eB)$ schemes. For scaled fourth order correlations (${\hat {\chi}}_{31}^{SQ}$, ${\hat {\chi}}_{22}^{QS}$) and scaled second order correlation ${\hat {\chi}}_{11}^{QS}$, IMC effect does not lead to qualitative difference but causes quantitative modifications. ${\hat {\chi}}_{31}^{SQ}$ and ${\hat {\chi}}_{22}^{QS}$ have lower values with IMC effect in $G(eB)$, $K(eB)$ and $T_0(eB)$ schemes than without IMC effect. ${\hat {\chi}}_{11}^{QS}$ increases faster (slower) with IMC effect in $T_0(eB)$ ($G(eB)$ and $K(eB)$) scheme(s) than that without IMC effect.

Figure~\ref{figxbqstpcimc} plots the scaled fourth order correlations of baryon number $B$, electric charge $Q$ and strangeness $S$ at pseudo-critical temperature of chiral restoration phase transition as functions of magnetic field with and without IMC effect. For ${\hat {\chi}}_{211}^{BQS}$ and ${\hat {\chi}}_{211}^{QBS}$, IMC effect does not lead to qualitative difference but reduces their values. ${\hat {\chi}}_{211}^{SBQ}$ increases with magnetic field in cases without IMC effect and with IMC effect in $T_0(eB)$ scheme, and its value is lowered by the IMC effect. ${\hat {\chi}}_{211}^{SBQ}$ decreases (firstly decreases and then increases) with magnetic field in case with IMC effect in $K(eB)$ ($G(eB)$) scheme.\\

\section{summary}
\label{summary}

The fourth order correlations $\chi^{BQ}_{31}$, $\chi^{QB}_{31}$, $\chi^{BQ}_{22}$, $\chi^{BS}_{31}$, $\chi^{SB}_{31}$, $\chi^{BS}_{22}$, $\chi^{QS}_{31}$, $\chi^{SQ}_{31}$, $\chi^{QS}_{22}$, $\chi^{BQS}_{211}$, $\chi^{QBS}_{211}$, $\chi^{SBQ}_{211}$ of baryon number $B$, electric charge $Q$ and strangeness $S$ at finite temperature, magnetic field and vanishing quark chemical potential are investigated in the frame of a three-flavor PNJL model. Together with our previous work~\cite{maoyangprd}, we shows a comprehensive results of second and fourth order correlations and fluctuations of conserved charges at finite magnetic field and temperature.

With vanishing and non-vanishing magnetic field, the correlations $\chi^{BQ}_{31}$, $\chi^{QB}_{31}$, $\chi^{BQ}_{22}$, $\chi^{QS}_{31}$ and $\chi^{QS}_{22}$ ($\chi^{SQ}_{31}$, $\chi^{BQS}_{211}$ and $-\chi^{SBQ}_{211}$) increase with temperature and show peak structure around the pseudo-critical temperature of chiral restoration and deconfinement phase transitions (around the pseudo-critical temperature of strange quark chiral restoration phase transition). $-\chi^{BS}_{31}$, $-\chi^{SB}_{31}$, $\chi^{BS}_{22}$ and $-\chi^{QBS}_{211}$ increase with temperature and have wide bumps around the pseudo-critical temperature of strange quark chiral restoration phase transition with vanishing magnetic field. Turning on magnetic field, another peak around the pseudo-critical temperature of chiral restoration and deconfinement phase transitions is observed.

Being analogous to studying the central-to-peripheral collision conditions with increasing magnetic field, and to understand the property of correlations along the phase transition line in $T-eB$ plane, the scaled correlations ${\hat {\chi}}_{mn}^{XY}=\frac{\chi_{mn}^{XY}({\text {eB}},T_{pc}^c({\text {eB}}))}{\chi_{mn}^{XY}({\text {eB}}=0,T_{pc}^c({\text {eB}}=0))}$ ($n+m=4,\ (n,m>0)$) and ${\hat {\chi}}_{211}^{XYZ}=\frac{\chi_{211}^{XYZ}({\text {eB}},T_{pc}^c({\text {eB}}))}{\chi_{211}^{XYZ}({\text {eB}}=0,T_{pc}^c({\text {eB}}=0))}$, with $X,\ Y,\ Z=B,\ Q,\ S,\ (X \neq Y \neq Z)$ at the pseudo-critical temperature $T_{pc}^c$ of chiral restoration phase transition are analyzed. Except ${\hat {\chi}}_{211}^{BQS}$, they increase with magnetic field, which is mainly induced by the increase of phase transition strength under external magnetic field. Among them, ${\hat {\chi}}_{31}^{BQ}$ is more sensitive to the magnetic field than other second order and fourth order correlations and fluctuations, and can be served as a better magnetometer of QCD.

The IMC effect is introduced into the PNJL model by the magnetic field dependent coupling between quarks $G(eB)$ or $K(eB)$, or magnetic field dependent interaction between quarks and Polyakov loop $T_0(eB)$. With and without IMC effect, the strength of chiral restoration and deconfinement phase transitions increases as the magnetic field grows. The main conclusion is robust when including IMC effect.\\

\noindent {\bf Acknowledgement:} Shijun Mao is supported by the National Natural Science Foundation of China under Grant No.12275204. Guoyun Shao is supported by the National Natural Science Foundation of China under Grant No.12475145 and Natural Science Basic Research Plan in Shaanxi Province of China (Program No. 2024JC-YBMS-018). Wenchao Zhang is supported by the National Natural Science Foundation of China under grant Nos. 11447024 and 11505108, and by the Natural Science Basic Research Plan in Shaanxi Province of China (Program No. 2023-JCYB-012).\\

%\section{appendix}
%
%
%%%%%%%%%%%%%%%%%%%%%%%%%%%%%%%%%%%%%%%%%%%%%%%%%%%%%%%%%%%%%%%%%%%%%%%%%%%%%%%%%%%%%%%%%%%%%%%%%%%%%%%%%%%%%%%%%%%%%%%%%%%%%%%%%%%%%%%%%%%%%%%%%%%%%
%\begin{figure}[htb]
%\begin{center}
%\includegraphics[width=6cm]{x11x11tpc.eps}
%\includegraphics[width=6cm]{x11x2tpc.eps}
%\end{center}
%\caption{}
%\label{figx2x2tpc}
%\end{figure}
%%%%%%%%%%%%%%%%%%%%%%%%%%%%%%%%%%%%%%%%%%%%%%%%%%%%%%%%%%%%%%%%%%%%%%%%
%
%%%%%%%%%%%%%%%%%%%%%%%%%%%%%%%%%%%%%%%%%%%%%%%%%%%%%%%%%%%%%%%%%%%%%%%%%%%%%%%%%%%%%%%%%%%%%%%%%%%%%%%%%%%%%%%%%%%%%%%%%%%%%%%%%%%%%%%%%%%%%%%%%%%%%
%\begin{figure*}[htb]
%\begin{center}
%\includegraphics[width=6cm]{x4btpc.eps}\includegraphics[width=6cm]{x2btpc.eps}\\
%\includegraphics[width=6cm]{x4qtpc.eps}\includegraphics[width=6cm]{x2qtpc.eps}\\
%\includegraphics[width=6cm]{x4stpc.eps}\includegraphics[width=6cm]{x2stpc.eps}
%\end{center}
%\caption{The scaled fourth and second order fluctuations of baryon number $B$, electric charge $Q$ and strangeness $S$ at pseudo-critical temperature of chiral restoration phase transition as functions of magnetic field with and without IMC effect. For all panels, the green, blue and red lines are the results in case of considering IMC effect in $K(eB)$, $T_0(eB)$ and $G(eB)$ scheme, respectively, and the black lines are the results without IMC effect.}
%\label{figx4tpcimc}
%\end{figure*}
%%%%%%%%%%%%%%%%%%%%%%%%%%%%%%%%%%%%%%%%%%%%%%%%%%%%%%%%%%%%%%%%%%%%%%%%

\end{document}